\documentclass[pra,showpacs,preprintnumbers,amsmath,amssymb,floatfix]{revtex4}
\usepackage{dcolumn}
\usepackage{bm}

\usepackage{amsmath}    % need for subequations
\usepackage{graphicx}   % for figures

\def\oper#1{\hat{\mathbf{#1}}} %zapis operatora
\begin{document}

%-- title and author  ---------------------------------------------
\title{Analytic pulse design for selective population transfer in many-level quantum systems: optimizing the pulse duration}
%Lines break automatically or can be forced with \\
\author{Duje Bonacci}
\affiliation{Bleiweissova 26, 10000 Zagreb, Croatia}
\email{duje@znanost.org}

\date{\today}

%--  abstract   --------------------------------------------------
\begin{abstract}
In the previous paper on this topic it was shown how, for a pulse
of arbitrary shape and duration, the drive frequency can be
analytically optimized to maximize the amplitude of the population
oscillations between the selected two levels in a general many
level quantum system. It was shown how the standard Rabi theory
can be extended beyond the simple two-level systems. Now, in order
to achieve the quickest and as complete as possible population
transfer between two pre-selected levels, driving pulse should be
tailored so that it produces only a single half-oscillation of the
population. In this paper, this second (and final) step towards
the controlled population transfer using modified (i.e. many level
system) Rabi oscillations is discussed. The results presented
herein can be regarded as an extension of the standard $\pi$-pulse
theory - also strictly valid only in the two level systems - to
the coherently driven population oscillations in general many
level systems.
\end{abstract}

\pacs{3.65.Sq}

\maketitle

%--  main text  ------------------------------------------------
\section{Introduction}

During the past 20 years a number of methods has been devised for
state selective preparation and manipulation of discrete-level
quantum systems
\cite{paramonov1983,chelkowski1990,kaluza1993,bergmann1998,rabitz2003}.
However, simple population oscillations, induced by a resonant
driving pulse have received negligible attention as a prospective
population manipulation method. This might be attributed to two
reasons. The first is that Rabi theory is based on the rotating
wave approximation (RWA), and all attempts to generalize it
without RWA (e.g. \cite{shariar2002.1,barata2000,fujii2003}) are
mathematically very involved. The second is that no attempt has
been made to analytically generalize the original Rabi theory
beyond the two-level systems.

In this paper an analytic extension of Rabi theory to transitions
in many-level systems is presented. The aim is to 'design' a
driving pulse of the form:
\begin{equation}
  F(t)= F_{\rm 0} \; m(t) \; \cos{(\omega(t) \; t)}
\label{pulse}
\end{equation}
by establishing analytical optimization relations between its
parameters: maximum pulse amplitude $F_{0}$, pulse envelope shape
$m(t)$, and time dependent carrier frequency $\omega(t)$. The goal
of this enterprize is twofold: the first is to achieve as complete
as possible transfer of population between two selected states of
the system; the second is to make this transfer as rapid as
possible. These two requirements, however, are conflicting:
population transfer can be accelerated by using a more intense
drive, but at the same time a stronger drive increases involvement
of remaining system levels in population dynamics hence
deteriorating population transfer between a selected pair of
levels.

In the previous paper on this topic \cite{bonacci2003.2} it was
shown how, for a pulse of arbitrary shape and duration, the drive
frequency can be analytically optimized to maximize the amplitude
of the population oscillations between the selected two levels in
a general many level quantum system. It was shown how the standard
Rabi theory can be extended beyond the simple two-level systems.
Now, in order to achieve the quickest and as complete as possible
population transfer between two pre-selected levels, driving pulse
should be tailored so that it produces only a single
half-oscillation of the population. In this paper, this second
(and final) step towards the controlled population transfer using
modified (i.e. many level system) Rabi oscillations is discussed.
The results presented herein can be regarded as an extension of
the standard $\pi$-pulse theory (see e.g. \cite{holthaus1994}) -
also strictly valid only in the two level systems - to the
coherently driven population oscillations in general many level
systems.

%-------------------------------------------------------------------
\section{Theoretical analysis}

All the calculations in this section are done in a system of units
in which $\hbar=1$.

\subsection{Calculation setup}
A quantum system with N discrete stationary levels with energies
$E_i \ (i=1,...,N)$ is considered. The system is driven by a time
dependent perturbation given in Eq. (\ref{pulse}). In the
interaction picture, the dynamics of the system obeys the
Schroedinger equation:
\begin{equation}
  \frac{d}{dt}\mathbf{a}(t)=-i \oper{V}(t) \mathbf{a}(t) ,
\label{schrodinger}
\end{equation}
where $\mathbf{a}(t)$ is a vector of time-dependent expansion
coefficients $a_1(t),..., a_N(t)$. The N$\times$N Matrix
$\oper{V}(t)$ describes interaction between the system and
perturbation. Explicitly, its elements are given by:
\begin{equation}
  V_{ij}(t) \equiv \frac{F_0 \mu_{ij}}{2} m(t)(e^{i s_{ij}
  (\omega(t)-\omega_{ij}) t}+e^{-i s_{ij} (\omega(t)+\omega_{ij}) \; t}).
\end{equation}
$\mu_{ij}$ is transition moment between the i-th and the j-th
levels induced by the perturbation. $s_{ij} \equiv sign(E_i-E_j)$
and $\omega_{ij} \equiv |E_i-E_j|$ are respectively the sign and
the magnitude of the resonant frequency for the transition between
the i-th and the j-th level.

The aim is to induce population transfer between two arbitrarily
selected levels, designated by $\alpha$ and $\beta$, directly
coupled by the perturbation (i.e. such that $\mu_{\alpha \beta}\ne
0$). To simplify equations, the time variable t is re-scaled to
$\tau$, with transformation between the two given by:
\begin{equation}
  d\tau \equiv \frac{F_0 \mu_{\alpha \beta}}{2} m(t) dt .
\label{definition tau}
\end{equation}
Then with following substitutions:
\begin{eqnarray}
  f_{ij}(\tau)&\equiv&s_{ij} \frac{2}{F_0 \mu_{\alpha \beta}} (\omega(t)-\omega_{ij}) \label{definition f} \\
  g_{ij}(\tau)&\equiv&s_{ij} \frac{2}{F_0 \mu_{\alpha \beta}} (\omega(t)+\omega_{ij}) \label{definition g} \\
  x(\tau)&\equiv& \frac{F_0 \mu_{\alpha \beta}}{2} t(\tau) \label{definition x} \\
  R_{ij} &\equiv& \frac{\mu_{ij}} {\mu_{\alpha \beta}} \label{definition R}
\end{eqnarray}
Eq. (\ref{schrodinger}) transforms into:
\begin{equation}
  \frac{d}{d \tau}\mathbf{a}(\tau)=-i \oper{W}(\tau)\mathbf{a}(\tau),
\label{n level}
\end{equation}
where:
\begin{equation}
  W_{ij}(\tau) \equiv R_{ij}(e^{i f_{ij}(\tau) x(\tau)}+e^{-i g_{ij}(\tau)x(\tau)}).
\label{wovi}
\end{equation}

Initial conditions for the problem of selective population
transfer comprise complete population initially (at $t=\tau=0$)
contained in only one of the selected levels, either $\alpha$ or
$\beta$. The other selected level, as well as all the remaining
N-2 'perturbing' levels of the system are unpopulated at this
time.

Population evolution $\Pi_i(t)$ of the i-th level is determined
from $\Pi_i(t)=|a_i(t)|^2$.
%----------------------------------------------------------------
\subsection{Rabi-like population transfer in a three level system}

It was demonstrated in the previous paper on this topic
\cite{bonacci2003.2} that the analytical extension of the Rabi
oscillations theory beyond two-level systems is anchored in the
analysis of the simplest of the 'many-level' systems - a three
level one. Hence, in this section the impact of the single
additional level on the population transfer period is discussed:
beyond the 'selected' levels $\alpha$ and $\beta$, the system now
contains one additional 'perturbing' level, designated with index
\textit{p}. The only requirements on the system internal structure
are that $\mu_{\alpha \beta}, \mu_{\beta p} \ne 0$ and
$\mu_{\alpha p}=0$. While the first two requirements are
necessary, the last one does not reduce the generality of the
final results to any significant extent and is introduced for
calculational convenience exclusively.

For the observed three-level system, the dynamical equation
(\ref{n level}) reduces to:

\begin{eqnarray}
\label{3 level}
  \frac{d}{d \tau}
  \begin{bmatrix}
  a_\alpha(\tau) \\
  a_\beta(\tau) \\
  a_p(\tau) \\
  \end{bmatrix}
  &=& \\
  -&i&
  \begin{bmatrix}
  0 & e^{i f_{\alpha \beta}(\tau) x(\tau)} + e^{-i g_{\alpha \beta}(\tau) x(\tau)} & 0 \\
  e^{-i f_{\alpha \beta}(\tau) x(\tau)} + e^{i g_{\alpha \beta}(\tau) x(\tau)} &  0
  & R_{\beta p}(e^{-i f_{\beta p}(\tau) x(\tau)} + e^{i g_{\beta p}(\tau) x(\tau)}) \\
  0 & R_{\beta p}(e^{i f_{\beta p}(\tau) x(\tau)} + e^{-i g_{\beta p}(\tau) x(\tau)}) & 0 \\
  \end{bmatrix}
  \begin{bmatrix}
  a_\alpha(\tau) \\
  a_\beta (\tau) \\
  a_p(\tau) \\
  \end{bmatrix} \nonumber
\end{eqnarray}

%----------------------------------------------------------------
\subsubsection{Recapitulation: minimizing the impact of the perturbing level}

As it was shown in \cite{bonacci2003.2}, Eq. (\ref{3 level}), when
the driving frequency is near the resonant value for the
transition $\alpha \leftrightarrow \beta$, the following
expression can be obtained for the population dynamics of level
$\textit{p}$:

\begin{equation}
\label{apert}
  a_p(\tau) \approx -  a_{\beta} (\tau)\Big( \sigma_{\beta p}
  \frac{ m(t(\tau)) }{1-\Delta_{\beta p}(\tau)} \Big) \; e^{ i f_{\beta
  p}(\tau) x(\tau)}
\end{equation}

where

\begin{eqnarray}
\label{delta i Delta}
 \sigma_{\beta p} \equiv \frac{ F_0 \mu_{\beta p}}{2(\omega_{\alpha \beta} - \omega_{\beta p})} \, \nonumber \\
 \Delta_{\beta p}(\tau) \equiv \frac {\omega(\tau)- \omega_{\alpha \beta}} {\omega_{\beta p} - \omega_{\alpha \beta}} \ .
\end{eqnarray}

Put in words, with the conditions mentioned, the dynamics of the
level \textit{p} parametrically depends on the dynamics of the
level $\beta$ to which it is coupled. The relation between the
amplitudes of the population oscillations for levels \textit{p}
and $\beta$ follows directly from the above expression:

\begin{eqnarray}
\label{beta and p}
  \Pi_{p} &\approx& \epsilon_{\beta p}(\tau) \Pi_{\beta}(\tau)
\end{eqnarray}

where:

\begin{equation}
\epsilon_{\beta p}(\tau) \equiv \Big(\sigma_{\beta p}
  \frac{m(t(\tau))}{1-\Delta_{\beta p}(\tau)}\Big) ^2 \ .
\end{equation}

Note that, as $\Delta_{\beta p}(\tau)$ is generally very small and
$|m(\tau)|\leq1$, that parameter $\sigma_{\beta p}$ actually
determines the effective strength of applied perturbation: if
$\sigma_{\beta p}^2<<1$, then dynamical impact of level p is
negligible and perturbation may be considered weak; if
$\sigma_{\beta p}^2\sim 1$, perturbation is very strong.

Further, the requirement of the minimization of the dynamical
impact of the perturbing level $\textit{p}$ on the transition
$\alpha \leftrightarrow \beta$ leads to the following equation for
the optimized dynamics of the ($\alpha$,$\beta$) subsystem:

\begin{equation}
  \label{beta dynamics}
  \frac {d }{d \tau}
  \begin{bmatrix}
  b_\alpha(\tau) \cr
  b_\beta(\tau)
  \end{bmatrix}
  = - i
  \begin{bmatrix}
  0 & 1 \cr
  1 & 0 \cr
  \end{bmatrix}
  \begin{bmatrix}
  b_\alpha(\tau) \cr
  b_\beta(\tau)
  \end{bmatrix}
\end{equation}

where the two-level state vector $\big(
b_\alpha(\tau),b_\beta(\tau)\big)$ is merely the unitary
transformed vector of the subsystem ($\alpha$,$\beta$):

\begin{equation}
\label{transformation}
  \begin{bmatrix}
  b_\alpha(\tau) \\
  b_\beta(\tau) \\
  \end{bmatrix}
  = e^{-i \oper{\Lambda} (\tau)}
  \begin{bmatrix}
  a_\alpha(\tau) \\
  a_\beta(\tau) \\
  \end{bmatrix}
\end{equation}

Note that the precise form of the real transformation matrix
$\hat{\mathbf{\Lambda}}(\tau)$ is irrelevant here as it has no
impact on the population dynamics.

The optimization procedure produces the analytic expression for
the chirp of the driving frequency, which in the lowest order of
approximation (suitable for all but the most intensive
perturbations) amounts:

\begin{equation}
\label{chirp}
  \omega(t)=
  \omega_{\alpha \beta} +
  (s_{\beta \alpha} s_{\beta p})( \omega_{\beta p}- \omega_{\alpha
  \beta})\frac{2 \omega_{\beta p} }{ \omega_{\alpha \beta} +\omega_{\beta p}}
  \sigma_{\beta p}^2 \frac{1}{t} \int_0^t \big( m(t') \big) ^2 dt'
\end{equation}

and from the Eq (\ref{beta dynamics}) it is found that the time
$\Theta$ required for a single population transfer between levels
$\alpha$ and $\beta$, determined from the fundamental relation of
the $\pi$-pulse theory:

\begin{equation}
\label{pi pulse theory}
  \int_0^\Theta d\tau = \frac{\pi}{2}
\end{equation}

equals (in units of $\tau$):
\begin{equation}
\label{period pi}
  \Theta=\frac{\pi}{2}
\end{equation}

As was shown in \cite{bonacci2003.2}, the 'exact' numerical
solution to the Eq. (\ref{3 level}) indeed maximizes the
population oscillations in the $\alpha-\beta$ subsystem. However,
as will be discussed below, the predicted value for the period of
the population oscillations (Eq. (\ref{period pi})) is somewhat
smaller than the correct one, with the discrepancy increasing with
the increasing population leak to the level \textit{p}. In the
following section this issue is resolved and the corrected
analytical expression for determination of the population transfer
(or oscillation) period is obtained.

%----------------------------------------------------------------
\subsubsection{Patching the population conservation of the total system}

To start the following discussion, notice that the optimized
solution for the population transfer between levels $\alpha$ and
$\beta$ (described by the Eq. (\ref{beta dynamics})) is
unfortunately too good to be true. Namely, its serious drawback
lays in the fact that the leak of the population from the
($\alpha$,$\beta$) subsystem into the perturbing level \textit{p}
goes by completely unnoticed!

Formally, the root of the problem hides in the fact that the
mathematical trick which enabled decoupling of the level
\textit{p} dynamics from the rest of the system (i.e. the step
between Eq.(18) and Eq.(19) in \cite{bonacci2003.2}) destroys the
unitarity of the full dynamical equation for the three level
system, Eq.(\ref{3 level}). The consequence is that the dynamical
equation for the $\alpha-\beta$ subsystem, Eq.(\ref{beta
dynamics}) itself claims to be unitary, keeping the population of
that subsystem fully conserved. This is clearly impossible, as
level \textit{p} does indeed capture some population - the exact
amount given by Eq.(\ref{beta and p}).

This malfunction caused by the decoupling procedure unfortunately
cannot be remedied within the decoupling procedure itself - the
patch has to be provided by an independent approach. To do this,
the following argument is used: since it is the equation for the
dynamics of level $\beta$ that changes due to the decoupling
procedure and consequently causes the breakdown of the population
conservation, it is only the dynamical equation for the level
$\beta$ that has to be modified; then, as the dynamics of the
level $\beta$ is governed by the elements in the lower row of the
dynamical matrix in Eq.(\ref{beta dynamics}), a particular ansatz
intervention precisely into these elements might help rectify the
overall population dynamics of the whole three level system.
Hence, the correction is sought in the following form:

\begin{equation}
  \label{two level corrected}
  \frac{d}{d\tau}
  \begin{bmatrix}
  b_{\alpha}(\tau) \\
  b_{\beta}(\tau) \\
  \end{bmatrix}
  = - i
  \begin{bmatrix}
  0 & 1 \\
  \zeta(\tau) & i \xi(\tau) \\
  \end{bmatrix}
    \begin{bmatrix}
  b_{\alpha}(\tau) \\
  b_{\beta}(\tau) \\
  \end{bmatrix}
\end{equation}

where $\zeta[\tau]$ and $\xi[\tau]$ are real non-negative
functions. Such an ansatz does not interfere with the
phase-fitting effect of the driving frequency optimization and Eq.
(\ref{chirp}) - forged by the decoupling procedure - which
maximizes the population oscillations in the $\alpha-\beta$
subsystem, is left unharmed. Instead, it merely enables
phase-independent modification of the \textbf{amplitudes} of
$\alpha$ and $\beta$ populations.

Expressing requirement of population conservation in the total
three-level system as:

\begin{equation}
d|b_{\alpha}(\tau)|^2+d|b_{\beta}(\tau)|^2+d|b_{p}(\tau)|^2=0 \ ,
\end{equation}

splitting the phase and amplitude contributions of the three wave
function projections on the three stationary states:

\begin{eqnarray}
 b_{\alpha}(\tau) &\equiv& B_{\alpha}(\tau) e^{i \phi_{\alpha}(\tau)}\  , \nonumber \\
 b_{\beta}(\tau) &\equiv& B_{\beta}(\tau) e^{i \phi_{\beta}(\tau)}\  ,  \\
 b_{p}(\tau) &\equiv& B_{p}(\tau) e^{i \phi_{p}(\tau)} \  \nonumber ,
\end{eqnarray}

and using the known relation between the populations of levels
$\beta$ and \textit{p}, Eq.(\ref{beta and p}), the following
result is obtained:

\begin{equation}
\label{cons condition}
 B_{\alpha}(\tau)\; dB_{\alpha} + (1+\epsilon_{\beta p}(\tau))
 B_{\beta}(\tau)\; dB_{\beta} + \frac{1}{2} B_{\beta}(\tau)^2 \; d\epsilon_{\beta p} =
 0\ .
\end{equation}

Now the corrected equation for the $\alpha-\beta$ subsystem,
Eq.(\ref{two level corrected}), can be used to eliminate
$dB_{\alpha}$ and $dB_{\beta}$ and introduce $\zeta[\tau]$ and
$\xi[\tau]$ in their stead:

\begin{eqnarray}
\label{eqn condition}
  dB_{\alpha}&=& -i B_{\beta} \; Im \big(
  e^{i(\phi_{\beta}(\tau)-\phi_{\alpha}(\tau))}\big)
  \; d{\tau} \ , \nonumber \\
  dB_{\beta}&=& -i \Big(\zeta(\tau) \;B_{\alpha}(\tau) \; Im \big( e^{-i(\phi_{\beta}(\tau)-\phi_{\alpha}(\tau))}\big) + \xi(\tau) \; B_{\beta}(\tau) Im \big(e^{-i\phi_{\beta}(\tau)}\big) \Big) \;
  d{\tau} \ .
\end{eqnarray}

Finally, taking together Eq.(\ref{cons condition}) and
Eq.(\ref{eqn condition}) it is found that:

\begin{equation}
\label{final condition}
 B_{\alpha}(\tau) \Big(\zeta(\tau)-\frac{1}{1+\epsilon_{\beta p}(\tau)}\Big)
 \ sin \big( \phi_{\beta}(\tau)-\phi_{\alpha}(\tau) \big)
+ B_{\beta}(\tau) \Big( \xi(\tau) + \frac{d }{d\tau}\ln
\big(1+\epsilon_{\beta p}(\tau)\big)^{\frac{1}{2}} \Big) sin \big(
\phi_{\beta}(\tau) \big)=0 .
\end{equation}

Since for population oscillations $B_{\beta}$ and $B_{\alpha}$ are
$180^o$ out of phase, this condition can be satisfied only if:

\begin{eqnarray}
\label{zeta xi}
 \zeta(\tau) &=& \frac{1}{1+\epsilon_{\beta p}(\tau)} \nonumber \ , \\
 \xi(\tau) &=& - \frac{d }{d\tau}\ln
\big(1+\epsilon_{\beta p}(\tau)\big)^{\frac{1}{2}} \ .
\end{eqnarray}

%----------------------------------------------------------------
\subsubsection{Population oscillation period modified}

Hence, the correct dynamical equation for the $\alpha-\beta$
subsystem which both maximizes the population oscillation
amplitudes of these two levels as well as properly conserves the
overall population of the three-level ($\alpha-\beta-p$) system
is:

\begin{equation}
  \label{two level final}
  \frac{d}{d\tau}
  \begin{bmatrix}
  b_{\alpha}(\tau) \\
  b_{\beta}(\tau) \\
  \end{bmatrix}
  = - i
  \begin{bmatrix}
  0 & 1 \\
  \frac{1}{1+\epsilon_{\beta p}(\tau)} & - i \frac{d }{d\tau}\ln
\big(1+\epsilon_{\beta p}(\tau)\big)^{\frac{1}{2}} \\
  \end{bmatrix}
    \begin{bmatrix}
  b_{\alpha}(\tau) \\
  b_{\beta}(\tau) \\
  \end{bmatrix} \ .
\end{equation}

A simple extension of this result to the general many-level system
(in which level $\alpha$ also has some perturbing levels - jointly
designated by q - attached to it) yields the total corrected
dynamical equation for such a system:

\begin{equation}
  \label{two level final}
  \frac{d}{d\tau}
  \begin{bmatrix}
  b_{\alpha}(\tau) \\
  b_{\beta}(\tau) \\
  \end{bmatrix}
  = - i
  \begin{bmatrix}
  - i \frac{d }{d\tau}\ln \big(1+\epsilon_{\alpha q}(\tau)\big)^{\frac{1}{2}} & \frac{1}{1+ \epsilon_{\alpha q}(\tau)} \\
  \frac{1}{1+\epsilon_{\beta p}(\tau)} & - i \frac{d }{d\tau}\ln
\big(1+\epsilon_{\beta p}(\tau)\big)^{\frac{1}{2}} \\
  \end{bmatrix}
    \begin{bmatrix}
  b_{\alpha}(\tau) \\
  b_{\beta}(\tau) \\
  \end{bmatrix} \ .
\end{equation}

Now to finalize the calculation of the corrected population
transfer period the following procedure is administered. First,
the time variable is transformed $\tau \rightarrow \varphi$
according to:

\begin{equation}
 \label{var transform}
 d\tau = \kappa(\varphi) d\varphi
\end{equation}

The goal of this variable transformation is to produce, in the new
time variable $\varphi$, the closed dynamical equations for
$b_\alpha(\varphi)$ and $b_\beta(\varphi)$ describing the dynamics
which is as close as possible to the simple harmonic oscillation.
Second, and to that end, the transformation Eq.(\ref{var
transform}) is introduced into Eq.(\ref{two level final}), the
resulting relation is differentiated with respect to $\varphi$ and
all but the lowest order terms in the small parameters
$\epsilon_{\alpha q}(\tau)$ and $\epsilon_{\beta p}(\tau)$ are
kept. Hence the following result is established:

\begin{eqnarray}
  \label{many level time correction}
   \frac{d^2 b_{\alpha}(\varphi)}{d\varphi^2} +
  \frac{1}{2} \frac{d}{d\varphi}\Big(\ln \frac{\big(1+\epsilon_{\alpha q}(\varphi)\big)^3
  \big(1+\epsilon_{\beta p}(\varphi)\big)}{\kappa^2} \Big)\;\frac{d b_{\alpha}(\varphi)}{d\varphi}
  +\frac{\kappa^2}{\big(1+\epsilon_{\alpha q}(\varphi)\big)
  \big(1+\epsilon_{\beta p}(\varphi)\big)}b_{\alpha}(\varphi)= 0
  \nonumber
  \\
  \frac{d^2 b_{\beta}(\varphi)}{d\varphi^2} +
  \frac{1}{2} \frac{d}{d\varphi}\Big(\ln \frac{\big(1+\epsilon_{\alpha q}(\varphi)\big)
  \big(1+\epsilon_{\beta p}(\varphi)\big)^3}{\kappa^2} \Big)\;\frac{d b_{\beta}(\varphi)}{d\varphi}
  +\frac{\kappa^2}{  \big(1+\epsilon_{\alpha q}(\varphi)\big)
  \big(1+\epsilon_{\beta p}(\varphi)\big)}b_{\beta}(\varphi)= 0
\end{eqnarray}

In the third and final step, the appropriate value of the free
parameter $\kappa(\varphi)$ is selected:

\begin{equation}
  \kappa(\varphi)^2 \equiv \big(1+\epsilon_{\alpha
  q}(\varphi)\big)
  \big(1+\epsilon_{\beta p}(\varphi)\big)
\end{equation}

which transforms Eq.(\ref{many level time correction}) into:

\begin{eqnarray}
\label{approx equations}
  \frac{d^2 b_{\alpha}(\varphi)}{d\varphi^2}+ \frac{d}{d\varphi}(\ln(1+\epsilon_{\alpha q}(\varphi)))\frac{db_{\alpha}(\varphi)}{d\varphi}+
  b_{\alpha}(\varphi)=0 \nonumber \\
  \frac{d^2 b_{\beta}(\varphi)}{d\varphi^2}+ \frac{d}{d\varphi}(\ln(1+\epsilon_{\beta p}(\varphi)))\frac{db_{\beta}(\varphi)}{d\varphi}+
  b_{\beta}(\varphi)=0
\end{eqnarray}

Both these equations are similar to the damped oscillator
equation. For negligible damping ($\epsilon_{\alpha q}(\varphi),
\epsilon_{\beta p}(\varphi)<<1$), they reduce to the harmonic
oscillator equations, in which case the population transfer time
 of $\pi/2$ is obtained in the variable $\varphi$. In the the
original time coordinate, $\tau$, the corrected time $\Theta$
required for a single population transfer between the levels
$\alpha$ and $\beta$ is then obtained from:

\begin{equation}
\label{transfer time corrected}
 \int_0^\Theta \frac{d\tau}{\sqrt{(1+\epsilon_{\alpha q}(\tau))(1+\epsilon_{\beta p}(\tau))}} =
 \frac{\pi}{2}
\end{equation}

Note the difference between this result, and the result in Eq.
(\ref{pi pulse theory}) obtained from the standard $\pi$-pulse
theory: in the lowest order approximation, the corrected
population transfer time is shorter then the one obtained from Eq.
(\ref{pi pulse theory}) by an order of
$\frac{1}{2}(\epsilon_{\alpha q}(\tau)+\epsilon_{\beta p}(\tau))$.

On the other hand, taking into consideration the damping factor in
the Eq.(\ref{approx equations}), approximating:

\begin{equation}
\label{corrected period}
  \frac{d}{d\varphi}\ln(1+\epsilon_{\alpha q, \beta
  p}(\varphi))\approx \frac{d}{d\varphi}\epsilon_{\alpha q, \beta
  p}(\varphi)
\end{equation}

and using the damped oscillator theory \cite{Kent1996}(p.246), an
increase in the damping (expressed as an increase in
$\epsilon_{\alpha q}(\tau)$ and $\epsilon_{\beta p}(\tau)$) leads
to the an increase in the population time transfer by an order of
$\frac{1}{2}\frac{d}{d\tau}\epsilon_{\alpha q, \beta p}^2(\tau)$.
As this correction is an order of magnitude smaller than the
correction obtained from Eq.(\ref{transfer time corrected}), it
can be neglected for all practical purposes. Furthermore, the
additional corrections to the transfer period due to the neglected
higher order elements in the Eq.(\ref{many level time correction})
are of the same order of magnitude as this correction due to the
first derivative component, and as they are impossible to obtain
analytically, the value of this whole second order correction for
the case of strong perturbations is somewhat shaky. However, this
is not an issue, as the whole optimization theory developed in
\cite{bonacci2003.2} - and on whose applicability the results of
the above analysis hinge - assumes rather modest perturbations,
and is not even expected to work properly for the extreme values.

%----------------------------------------------------------------
\section{Numerical simulations}

\begin{figure}
  \includegraphics[width=8cm]{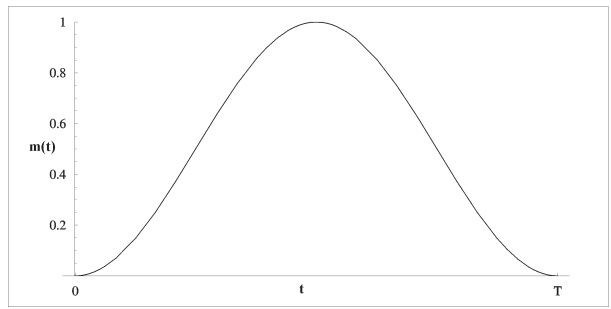}\\
\caption{In all of the numerical examples the same pulse form was
used - $m(t)= \sin(\Omega t)^2$, but with different values of
maximum intensity parameter ($F_0$) and different total pulse
duration, $T$. The respective values are quoted in each particular
example.}
  \label{fig1}
\end{figure}

In this section, numerical simulations of system dynamics for
unoptimized and fully optimized driving pulse of the form of Eq.
(\ref{pulse}) are presented and compared. Here, unoptimized
driving pulse is the one with driving frequency equal to the pure
resonant frequency between the two levels selected for population
transfer ($\omega(t)=\omega_{\alpha \beta}$) and with pulse
duration $T$ determined according to the standard $\pi$-pulse
theory relation, Eq.(\ref{pi pulse theory}). On the other hand,
the parameters of the fully optimized pulse are determined from
Eq.(\ref{chirp}) and Eq.(\ref{transfer time corrected}).

In all cases, a three-level system is considered, with the
following system parameters ($a.u.\equiv atomic \;units$):
$\omega_{\beta \alpha}=0.017671 \;a.u.$, $s_{\beta \alpha}=1$,
$\mu_{\beta \alpha}=0.073 \;a.u.$; $\omega_{\beta p}=0.017611
\;a.u.$, $s_{\beta p}=-1$, $\mu_{\beta p}=0.098\; a.u.$. These
system parameters correspond to the three ro-vibrational levels of
the HF molecule in the ground electronic state: $\alpha \equiv
(v=0,j=2,m=0)$, $\beta \equiv (v=1,j=1,m=0)$, $p \equiv
(v=2,j=2,m=0)$. The pulse shape in all of the examples is $m(t)=
\sin(\Omega t)^2$ as shown in Fig 1.

\subsection{Population oscillations}

\begin{figure}
  \includegraphics[width=12cm]{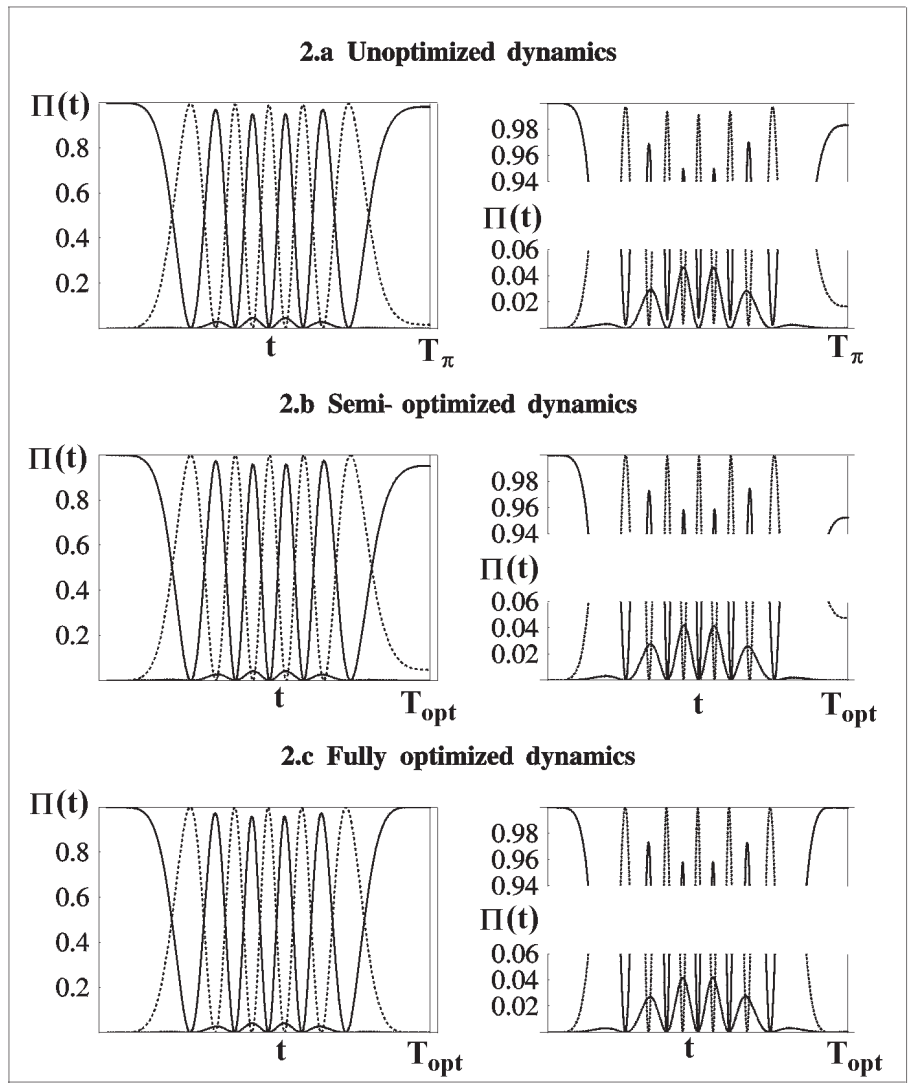}\\
\caption{Significance of the total analytical correction (in
driving frequency and total pulse duration) for the dynamics of a
mildly disturbed system. In all plots, major oscillations are the
populations of the two targeted levels ($\alpha$ and $\beta$)
whereas the minor oscillations are the population of the
perturbing level ($p$). For the value of perturbation strength
parameter $\sigma_{\beta p}^2=0.05$, the loss of the final
unoptimized population transfer amplitude amounts about 2\%.
Optimization reduces this loss to below 0.05\%. The difference
between the pulse duration obtained from standard $\pi$-pulse
theory and the optimized value is 1.6\%. Right hand-side plots
present the details from the left hand-side plots.}
  \label{fig2}
\end{figure}

As was demonstrated in \cite{bonacci2003.2}, frequency
optimization minimizes the impact of the perturbing levels on the
\textit{amplitude} of the population oscillations. In this
subsection, the necessity of the inclusion of additional
correction Eq.(\ref{transfer time corrected}) for the
\textit{population transfer time} - alongside the correction for
the driving frequency - will be demonstrated. Also, the validity
and the limitations of the analytically obtained expression for
this correction will be discussed.

\subsubsection{Legitimate perturbation}
Fig.2 presents the dynamics of the system subjected to the
external drive of limiting intensity, $\sigma_{\beta p}^2=0.05$,
corresponding to the $F_0=2.80534 \ast 10^{-4} a.u.$. It is just
strong enough to noticeably (albeit not significantly) distort the
pure resonant oscillations, but at the same time weak enough so
that the theory developed in \cite{bonacci2003.2} and further in
this paper provides the full and precise quantitative corrections.

Pulse duration determined according to the standard $\pi$-pulse
theory expression, Eq.(\ref{pi pulse theory}) is $T_{\pi}=3077832
a.u.$, whereas the optimized one, obtained from Eq.(\ref{transfer
time corrected}) is $T_{opt}=3126029 a.u.$. The pulse is aimed at
producing five complete population oscillations.

Three cases of dynamics are presented: Fig. 2.a shows the
unoptimized dynamics; Fig 2.b shows the 'semi-optimized' dynamics,
with optimized driving frequency, but unoptimized population
transfer period; finally, Fig. 2.c shows the fully optimized
dynamics. Observe that in the unoptimized case, the population
oscillations end somewhat short of the complete cycle, and the
initially populated level never achieves complete depopulation.
Optimizing only the driving frequency does indeed maximize the
population oscillations by inducing the complete depopulation of
the initially populated level during oscillations, but at the same
time the final population oscillation stops even further from the
full cycle than in the unoptimized case. Finally, introducing the
population transfer period correction alongside the driving
frequency correction yields the required result: complete cycle of
maximized population oscillations.

\subsubsection{Strong perturbation}
Increasing the driving perturbation intensity to somewhat greater
value, $\sigma_{\beta p}^2=0.25$ ($F_0=6.11409 \ast 10^{-4} \
a.u.$.), the limitations of the analytical theory clearly emerge.
This is shown in Fig. 3: Fig. 3.a - Fig. 3.c respectively show the
unoptimized, analytically optimized (according to Eq.(\ref{chirp})
and Eq.(\ref{transfer time corrected})) and 'manually optimized'
dynamics. The corresponding pulse duration times, aimed at
producing three complete population oscillations, are
$T_{\pi}=847324 \ a.u.$, $T_{opt}=901075 \ a.u.$ and
$T_{man}=884300 \ a.u.$.

Notice that in the analytically optimized case, Fig. 3.b, the
initially populated level still fully depopulates, which indicates
that even for this rather strong perturbation, the frequency
correction Eq.(\ref{chirp}) still stands strong. However, the
corrected period, although closer to the correct value than in the
unoptimized case, is still somewhat removed from the correct
value. Unfortunately, this 'optimization error' cannot be remedied
analytically. Remember that the analytical result
Eq.(\ref{transfer time corrected}) is obtained using only the
first order approximation (Eq.(\ref{many level time correction}))
to the full dynamical equations Eq.(\ref{two level final}). With
perturbation as strong as in this case, the dynamical impact of
the neglected elements of that equation begin to show. However, as
shown in Fig. 3.c, the full cycle of oscillations can still be
produced, but this additional correction to the pulse duration had
to be found by hand, using the trial and error method.

\begin{figure}
  \includegraphics[width=12cm]{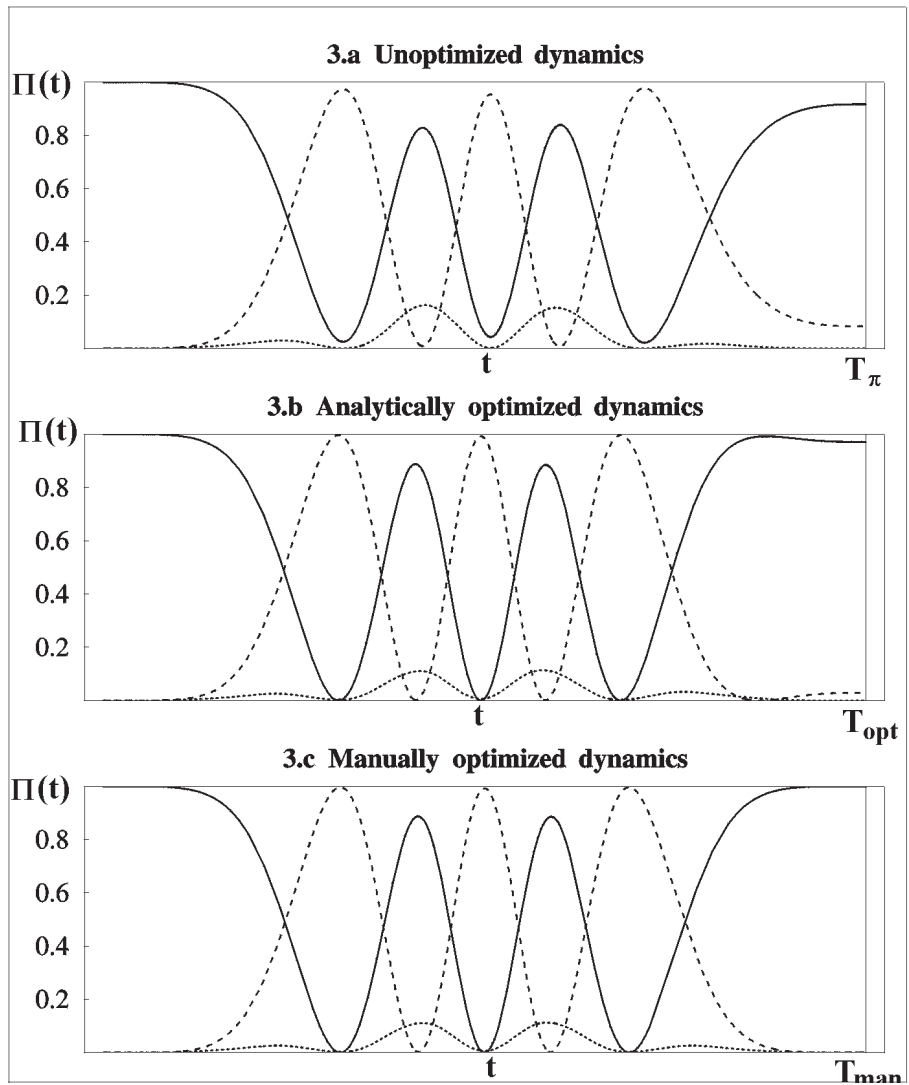}\\
\caption{Limitations of the analytical optimization theory. Again,
major oscillations are the population of the two targeted levels
whereas the minor oscillations are the population of the
perturbing level. For the value of perturbation strength parameter
is now $\sigma_{\beta p}^2=0.25$, which is just beyond the
limiting value for the full applicability of the presented
optimization procedure. Although the analytical correction
improves the final population transfer from 92\% to 97\% (with
pulse duration correction of 6\%), the theory presented in this
paper cannot account for an additional 1.6\% correction in the
duration of the pulse that further increases the final population
transfer to over 99.99\%.}
  \label{fig3}
\end{figure}

\subsection{Population transfer}
The two final examples demonstrate the application of the
developed optimization theory to the most interesting dynamical
case regarding the coherent control: that of the single population
transfer between the two targeted levels $\alpha$ and $\beta$. As
the validity and the limitations of the whole theory were already
explored in the previous two examples, the following examples will
only demonstrate the improvements to the population transfer that
can be produced using the above results.

\subsubsection{Legitimate perturbation}
Again as in the previous section, the first example (Fig. 4)
presents the dynamics of the system subjected to the external
drive of limiting intensity. In this case, the perturbation
strength parameter amounts $\sigma_{\beta p}^2=0.1$, corresponding
to the $F_0=4.07606 \ast 10^{-4} \ a.u.$. Calculated population
transfer times are $T_{\pi}=211831 \ a.u.$ and $T_{opt}=218483 \
a.u.$.

The unoptimized (dotted line) and the optimized dynamics (solid
line) are plotted on the same graph to facilitate the comparison
between the two. Only the dynamics of the two target levels is
shown - the plot of the perturbing level's ($p$) dynamics is
omitted for the sake of clarity of the overall graph. Although the
loss of the population transfer in the unoptimized case is not
great, it nevertheless is noticeable. On the other hand,
introducing the corrections for pulse frequency and pulse duration
clearly improves the population transfer bringing it very close to
100\%.

\begin{figure}
  \includegraphics[width=12cm]{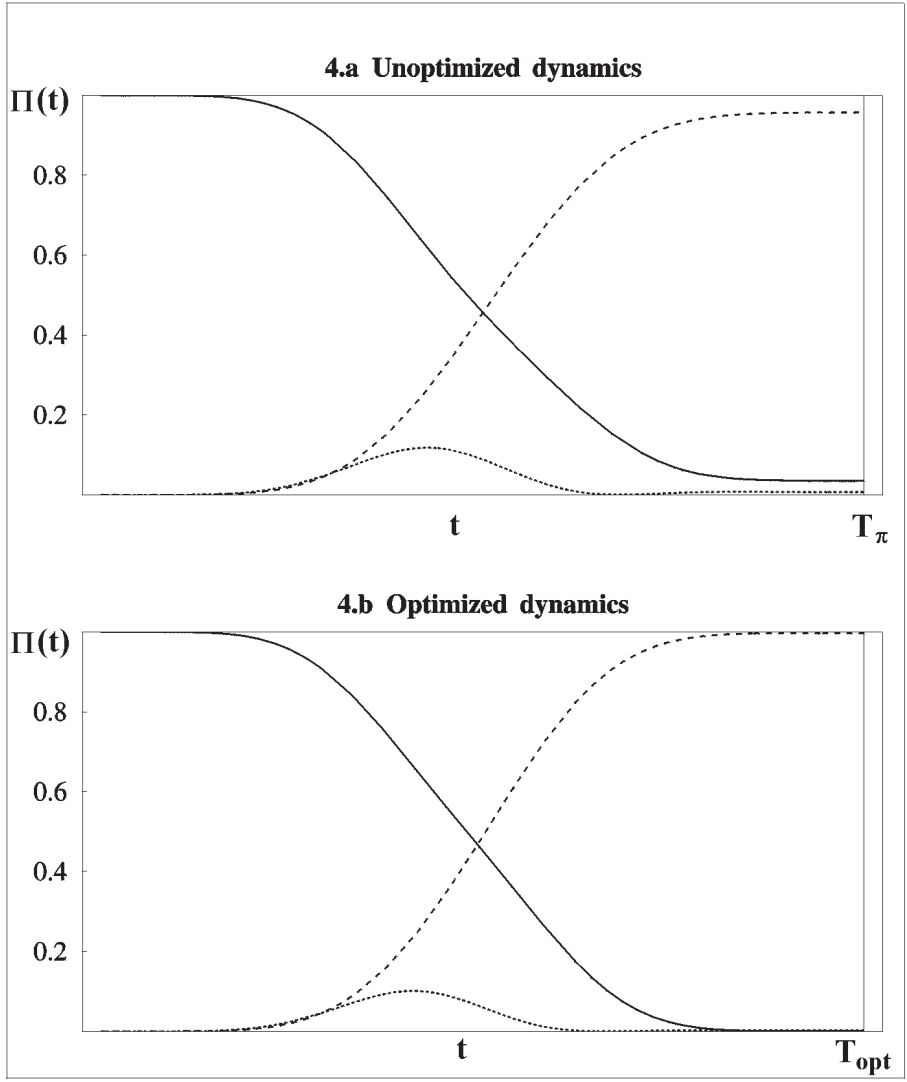}\\
\caption{Applicability of the analytical optimization procedure to
the maximization of the population transfer. Yet again, major
oscillations are the population of the two targeted levels whereas
the minor oscillations are the population of the perturbing level.
For this limiting value of perturbation strength parameter of
$\sigma_{\beta p}^2=0.10$, the optimization almost completely
eradicates the loss of the population transfer due to the
dynamical impact of the perturbing level, increasing the
population transfer from 96\% to 99.7\%. Optimized pulse lasts 3\%
longer than the one obtained from the standard $\pi$-pulse
theory.}
  \label{fig4}
\end{figure}

\subsubsection{Extreme perturbation}
The final example - presented in Fig. 5. - is qualitative, rather
than quantitative, but even as such it is quite indicative of the
overall usefulness of the whole optimization theory. The
perturbation is now extreme, with strength parameter
$\sigma_{\beta p}^2=1$ corresponding to the $F_0=1.22282 \ast
10^{-3}\ a.u.$. Calculated population transfer times are
$T_{\pi}=70610 \ a.u.$ and $T_{opt}=82816 \ a.u.$.

Again, the unoptimized and the optimized dynamics are plotted on
the same graph. Although optimization now clearly does not lead to
the complete population transfer, the improvement from the
unoptimized dynamics is significant demonstrating that even for
this perturbation intensity the developed optimization theory
qualitatively works quite nicely.

\begin{figure}
  \includegraphics[width=12cm]{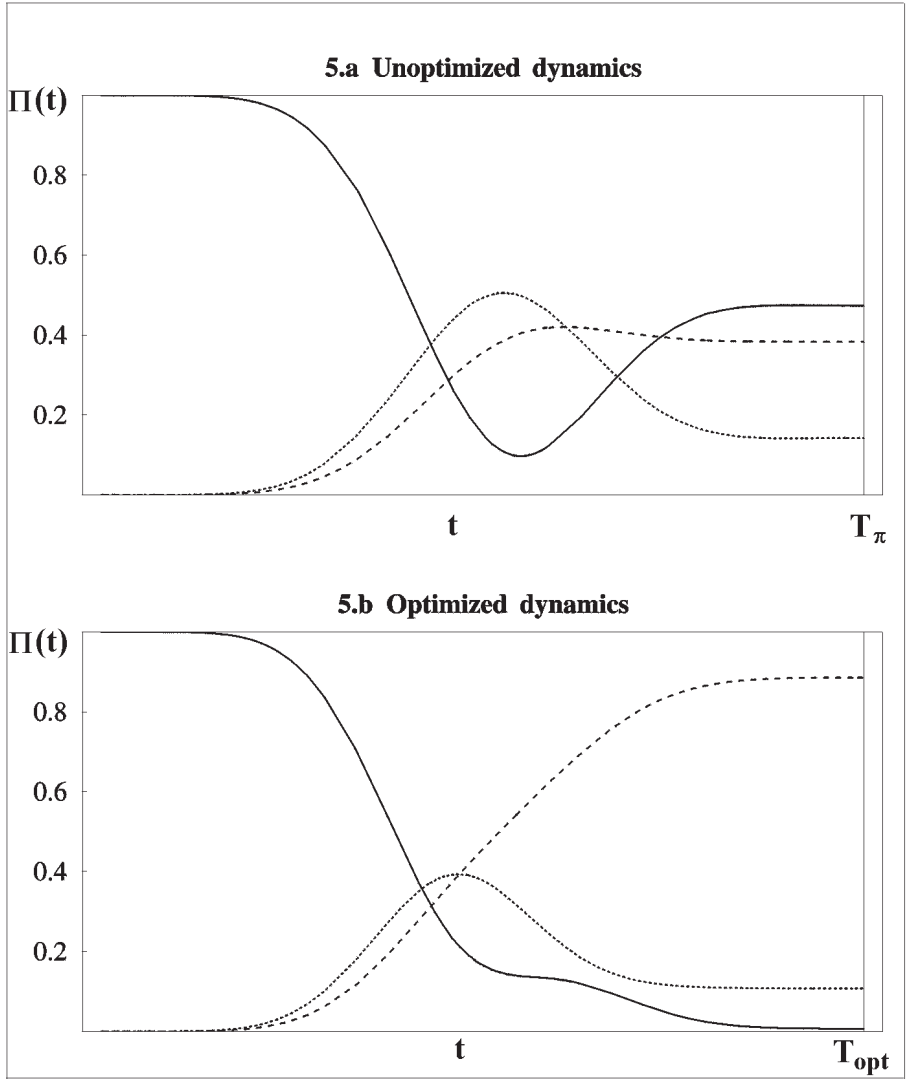}\\
\caption{Breakdown of the quantitative optimization, but
qualitatively the theory is still applicable. Half way through the
pulse, in the unoptimized case now the population of the
perturbing level surpasses the population of the initially
unpopulated level. Introduction of the optimized driving frequency
and pulse duration significantly - although not fully - rectifies
the population transfer from below 40\% to almost 90\%. Optimized
pulse is now almost 20\% longer then the one obtained from the
ordinary $\pi$-pulse theory.} \label{fig5}
\end{figure}

%----------------------------------------------------------------
\section{Conclusion}

The aim of research that led to this paper was to explore the
possibility of using 'old fashioned' and rather simple phenomenon
of Rabi oscillations for the controlled manipulation of the
population in general many level system. This paper rounds up the
topic of analytical optimization of pulse parameters (frequency
chirp and pulse duration), opened in the author's previous work
(\cite{bonacci2003.2}) that would lead to maximizing the
population transfer between two targeted levels of the system. The
theory developed provides the exact quantitative predictions of to
what extent the dynamical impact of the remainder of the many
level system (beyond the two levels selected for the population
transfer) begins to interfere with the targeted population
transfer. It also provides the closed (albeit recursive)
analytical expressions for the optimization of pulse parameters.

Although the major correction to the population transfer is
achieved by optimizing the driving pulse's frequency chirp (given
in \cite{bonacci2003.2}), this paper provides the additional fine
tuning by establishing similarly simple analytical expression for
the determination of the optimal pulse duration. It demonstrates
that the standard formula of the $\pi$-pulse theory, Eq. (\ref{pi
pulse theory}) begins to fail as the perturbation increases to and
beyond the well defined limiting value. It also provides some
remedy to this failure.

The whole theory presented in \cite{bonacci2003.2} and this paper
deals with only single laser pulse driving one particular
transition in the many level system. The further research
currently under way considers the possibility of applying a number
of distinct but simultaneous optimized pulses to drive the
population through the chain of transitions through the system,
hence producing as clean as possible transfer between the two
levels not coupled by the single photon transition. The
preliminary results indicate that an analytical optimization
formula can be developed even for such a case.

%%--  bibliography  ---------------------------------------------

%%---------------------------------------------------------------
%%---------------------------------------------------------------
\end{document}